%% file: paper.tex
 \pdfoutput=1 
\documentclass{egpubl}
\usepackage{egsr2023}
 
\SpecialIssuePaper         

\CGFStandardLicense

\usepackage[T1]{fontenc}
\usepackage{dfadobe}  

\usepackage{cite}  
\BibtexOrBiblatex
\electronicVersion
\PrintedOrElectronic

\ifpdf \usepackage[pdftex]{graphicx} \pdfcompresslevel=9
\else \usepackage[dvips]{graphicx} \fi

\usepackage{egweblnk} 

\usepackage{booktabs} 
\usepackage[export]{adjustbox} 

\usepackage{tikz}
\usetikzlibrary{positioning,calc,shadows,shadows.blur}
\usepackage{pgfplots}
\pgfplotsset{compat=newest}
\usepgfplotslibrary{fillbetween}

\usepackage{subcaption}
\usepackage{xcolor}
\usepackage{nicefrac}

\usepackage{soul}
\usepackage[normalem]{ulem} 

\usepackage{amsmath}
\usepackage{amssymb}
\usepackage[boxruled,lined]{algorithm2e} 
\SetCommentSty{emph}

\newcommand{\change}[2]{%
\if #1\empty%
\else%
{\color{red}%
\sout{#1}%
}%
\fi%
{\color{green!75!black}%
#2%
}%
}

\renewcommand{\change}[2]{%
#2%
}

\ccsdesc[500]{Computing methodologies~Rendering}

\title[One-to-Many Spectral Upsampling of Reflectances and Transmittances]{%
One-to-Many Spectral Upsampling \\ of Reflectances and Transmittances %
}
\author[L. Belcour, P. Barla \& G. Guennebaud]
{
\parbox{\textwidth}{\centering L. Belcour$^{1}$, P. Barla$^{2}$ and G. Guennebaud$^{2}$} \\
{\parbox{\textwidth}{ \centering
    $^1$Intel Corporation\hspace{0.5cm}
    $^2$INRIA, Bordeaux
}}
}

\begin{document}

\teaser{
    \centering
   {\footnotesize
	\begin{tikzpicture}[]
        	       \definecolor{tab_blue}{rgb}{0.12156862745098039,0.4666666666666667,0.7058823529411765}
        \definecolor{tab_orange}{rgb}{1.0,0.4980392156862745,0.054901960784313725}
		\definecolor{tab_green}{rgb}{0.17254901960784313,0.6274509803921569,0.17254901960784313}
        \begin{scope}
            \node[]               (A) {\includegraphics[height=5cm]{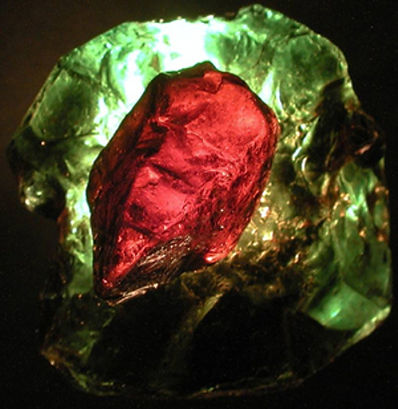}};
            \begin{scope}[yshift=-1.7cm]
                \node (spec_usanbara) {\includegraphics[width=4.0cm]{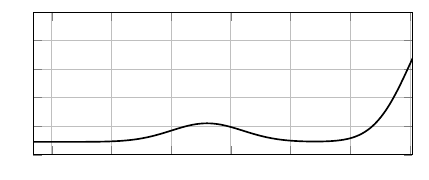}};
          \end{scope}
        \end{scope}
        \begin{scope}[xshift=5.80cm]
          \begin{scope}
              \clip (0,-2.5) rectangle (2.5,2.5);
              \node (B) {\includegraphics[height=5cm]{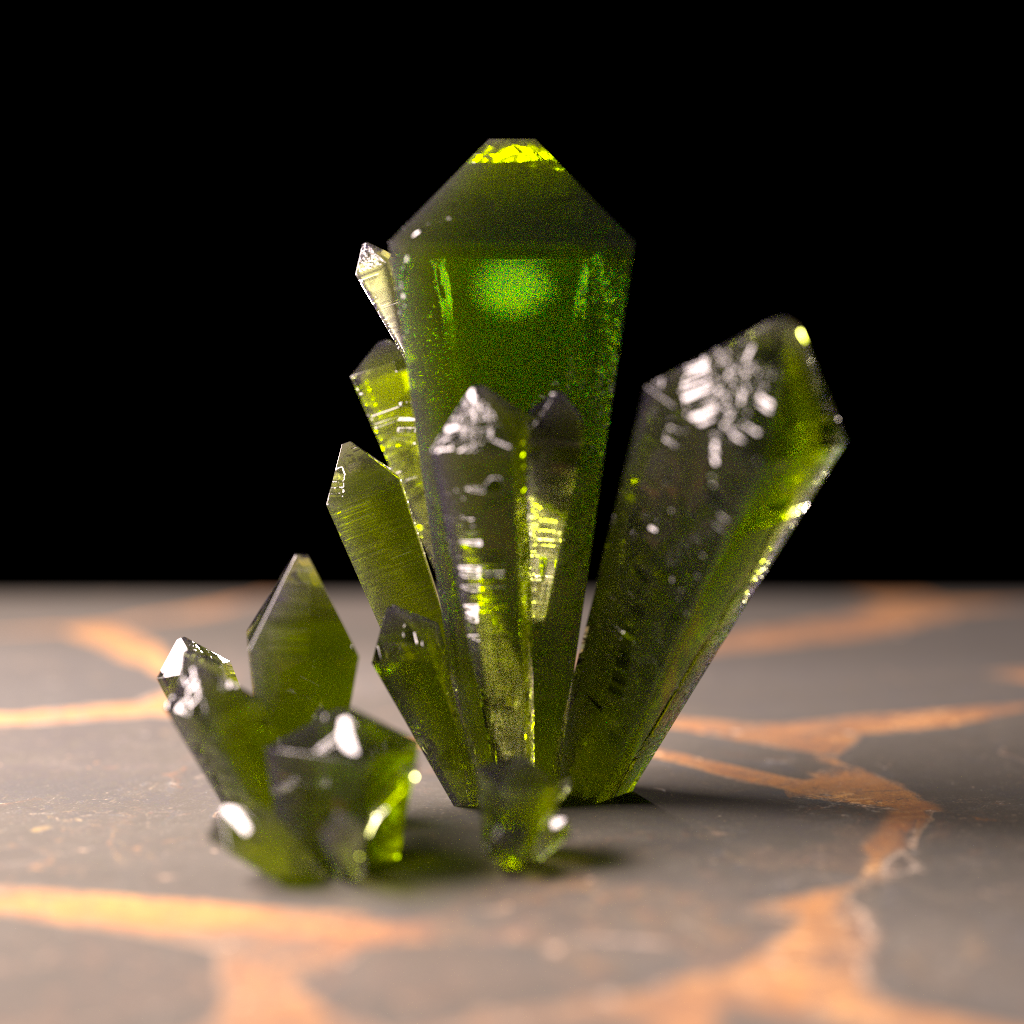}};
              \draw (1.25, 2.3) node[text=tab_green] (Ct) {{\tiny \textbf{Spectrum B}}};
              \node[text=white, below=-3pt of Ct] {{\tiny {no color effect}}};
          \end{scope}
          \begin{scope}
              \clip (-2.5,-2.5) rectangle (0,2.5);
              \node (C) {\includegraphics[height=5cm]{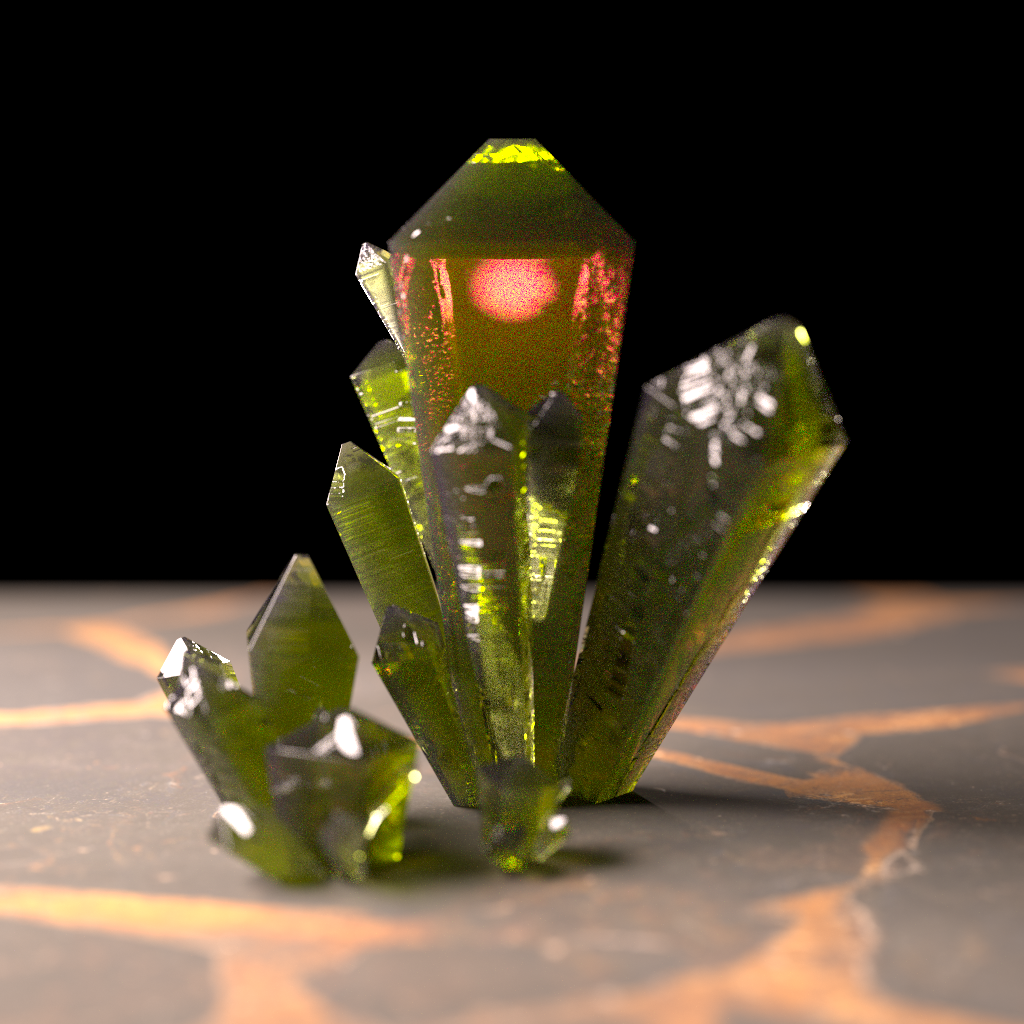}};
              \draw (-1.25, 2.3) node[text=tab_orange] (Ct) {{\tiny \textbf{Spectrum A}}};
              \node[text=white, below=-3pt of Ct] {{\tiny {Usambara effect}}};
          \end{scope}
          \draw ( 0.0,-2.5) rectangle (2.5,2.5); 
          \draw (-2.5,-2.5) rectangle (0.00,2.5); 
          \begin{scope}[yshift=-1.7cm]
            \node (spec) {\includegraphics[width=4.0cm]{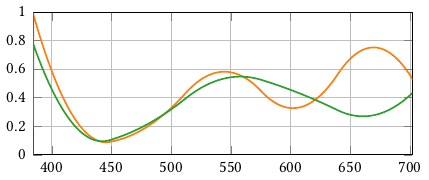}};
          \end{scope}
       \end{scope}
       \begin{scope}[xshift=11.6cm, yshift=-0.175cm]
       \node (D) {\includegraphics[height=5.3cm,width=5.3cm]{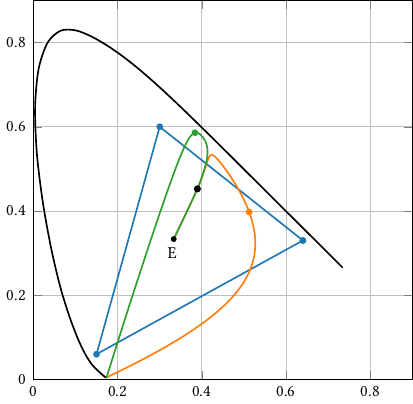}};
       \node at (0,0.35) {\textcolor{black}{\tiny $\mathbf{d=1}$}};
       \node at (1,-0.1) { \textcolor{tab_orange}{\tiny $\mathbf{d=10}$}};
       \node at (-0.1, 1.1) { \textcolor{tab_green}{\tiny $\mathbf{d=10}$}};
       \node at (-1.5, 0.0) { \textcolor{tab_blue}{\tiny\textbf{sRGB gamut}}};
       \end{scope}
       %
       \node[below=5pt of A] (tA) {\textbf{ a) Photograph of tourmaline}};
       \node[below=5pt of B] (tB) {\textbf{b) One-to-many spectral upsampling}};
       \node[below=-4pt of D] (tD) {\textbf{c) Corresponding chromaticities w.r.t. depth}};
	\end{tikzpicture}
    \caption{
        Surprising color changes may occur due to the path length travelled by light through some materials. 
        This is shown in (a) with the Usambara effect observed in tourmaline: the same gem appears green or red depending on path length, due to a specific transmission spectrum (see inset).
        We introduce a method for finding such non-generic spectra. 
        This requires to perform one-to-many spectral upsampling (b), which generates multiple spectra with a unique controllable color at a given path length, but varying colors at further lengths.
        The chromaticity of each spectrum as a function of path lengths is shown in (c).
        \label{fig:teaser}
    \vspace{5pt}
    }
   }
}

\maketitle

\begin{abstract}
Spectral rendering is essential for the production of physically-plausible synthetic images, but requires to introduce several changes in the content generation pipeline. 
In particular, the authoring of spectral material properties (e.g., albedo maps, indices of refraction, transmittance coefficients) raises new problems.
While a large panel of {}{computer graphics} methods exists to upsample a RGB color to a spectrum, they all provide a \emph{one-to-one} mapping. 
This limits the ability to control interesting color changes such as the Usambara effect or metameric spectra. 
In this work, we introduce a \emph{one-to-many} mapping in which we show how we can explore the set of all spectra reproducing a given input color. 
We apply this method to different colour changing effects such as vathochromism -- the change of color with depth, and metamerism.
\end{abstract}

\section{Introduction}

Spectral rendering has been increasingly used in recent years, due to raising expectations in photo-realism in cinematography~\cite{fascione2018manuka}, or to applications that require predictive results such as in architecture.
However, the generation of spectral material properties presents a challenge for artists and designers~\cite{weidlich2022practical}.
To ease the edition of reflectance and transmittance spectra, several spectral upsampling methods have been introduced {}{in computer graphics}~\cite{peercy1993linear,smits1999rgb,otsu2018reproducing,mallett2019spectral,jakob2019low,guarnera2022spectral,jendersie2021fast}. 
They produce spectra from colors, ensuring that physical bounds  are achieved (e.g., reflectances must lie in the $[0,1]$ range).
All of these methods are restricted to produce a \emph{one-to-one} conversion: one RGB triplet converts to a single spectrum.

This limitation restricts the possibilities that spectral rendering offers.
Indeed, unlike RGB materials, spectral materials offer the possiblity to produce
subtle color effects, such as metamerism -- a change of color due to different illuminants~\cite{weidlich2021spectral}.
{}{In the optics community, Metameric blacks~\cite{wyszecki1958evaluation,finlayson2005metamer} have been introduced to explore the space of metamers. 
In this approach, metameric spectra achieving a given desired color are sampled from a null-space in the target color space. 
Unfortunately, this requires to manipulate physical constraints in a high dimensional null-space, which significantly complicates artistic control.}
{}{Most importantly, this null-space approach is not easily adapted to deal with non-linear color changes, such as those observed in the Usambara effect (see Figure~\ref{fig:teaser}~(a)) -- a surprising change of color due to the path length travelled by light in tourmaline gems.}

In this paper, we introduce a novel \emph{one-to-many} approach that enables artists to design non-generic spectra with controlled color effects.
The key idea is to build reflectance or transmittance spectra using a small set of basis functions forming a partition of unity (PU), and to express them in chromaticity space.
{}{We primarily target non-linear effects: our }representation allows us to find many spectra that achieve a same target chromaticity at a unit optical depth, while providing control over the chromaticities at further depths.
This is shown in Figure~\ref{fig:teaser}(b,c) for a pair of spectra.
    
{}{A key observation on which we elaborate in Section~\ref{sec:PU-geom} is that a PU spectral representation is linked to generalized barycentric coordinates in chromaticity space.}
As demonstrated in Section~\ref{sec:PU-explore}, exploring all the possible barycentric coordinates reconstructing the same target chromaticity point is equivalent to exploring the space of all spectra producing the same chromaticity when integrated with respect to color matching functions of the human visual system.
We then use this geometric analogy for the design of the PU basis in Section~\ref{sec:PU-design}, where we show how to strike a balance between spectral smoothness and color expressivity.

With our one-to-many spectral upsampling approach, we are able to generalize the Usambara effect to any non-generic spectrum that exhibits changes of color with depth, which we {}{suggest to} call \emph{vathochromism}, derived from ancient Greek \textit{vathos} (depth) and \textit{chroma} (colour).\footnote{We reserve the usage of the term ``Usambara effect'' for the typical green-to-red color shift observed in Tourmaline gemstones.}
We show in Section~\ref{sec:apps-vatho} how to build a parametric system in which a user can pick spectra with specific constraints (such as reproducing two given chroma for different optical depths).
The same representation also provides control over metamerism, as shown in Section~\ref{sec:apps-meta}.
{}{We further discuss the differences with Metameric Blacks in Section~\ref{sec:discuss}.}

\input{tex/sec_previous_works.tex}

\input{tex/sec_partition_of_unity.tex}

\input{tex/sec_nullspace_sampling.tex}

\input{tex/sec_basis_design.tex}

\input{tex/sec_applications.tex}

\input{tex/sec_conclusion.tex}

\bibliographystyle{eg-alpha-doi}  
\bibliography{bibliography}

\end{document}

%% file: tex/sec_previous_works.tex
\section{Previous Work}
\label{sec:prev}

\subsection{Color changes in Nature}

Many different kinds of natural materials exhibit changes of color, depending on the angle of view (goniochromism), on temperature (thermochromism) or exposition to light (photochromism) for instance.
In all those cases, the reflected or transmitted spectrum is itself changed either due to an alteration of the material itself, or to viewing conditions.
In this paper, we are instead interested in materials that exhibit color changes despite the fact that their spectral reflectance or transmittance does \emph{not} change.

\paragraph*{Metamerism.} 
A common example of such materials are those that change color with a change of illumination, called metamers.
Two metameric materials can look the same under one illuminant, but will differ when lit by another illuminant.
This is explained by the fact that the human visual system integrates the product of light and material on photo-receptors, which is a many-to-one mapping.
Another related example of the impact of the illuminant on the appearance of objects is the \textit{Alexandrite effect}~\cite{gubelin1982gemstones}. 
Alexandrite gems are known to change from green when lit by sunlight to red when lit by candle light. 
This particular effect has been used in computer graphics by Bergner et al.~\cite{bergner2009tool} for visualization purposes.

\paragraph*{Usambara effect.} The Usambara effect was first described for a particular tourmaline found in the Umba valley in Tanzania~\cite{liu1999}. 
It was described as a change of color (from green to red) with an increase of the optical depth of the material. It was later found that other materials (such as topaz and amber) depicted such behaviour~\cite{bonventi2012color,liu2014}. 
In this work, we use the term \textit{vathochromism} for such changes of color with depth.

{
\subsection{Computing Metamers in Optics}

\paragraph*{Metameric blacks}
One way to generate a pair of metameric spectra is to add to a first spectrum a spectral curve that corresponds to a black color (i.e., a zero triplet) in the target color space~\cite{wyszecki1958evaluation}. 
From the point of view of linear algebra, a metamer is then a point in the null-space of the color space matrix~\cite{takahama1972new,cohen1982metameric}. 
While this formalism permits the generation of arbitrary metamers, it requires to track hard constraints: a reflectance spectrum must take its values in the $[0,1]$ range. 
With finely-discretized spectra, those constraints generate a convex hull of valid spectra in a high dimensional space. 
Alternate methods can avoid this dimensionality issue by using a blending of measured spectra~\cite{finlayson2005metamer,schmitt1976method}. 
However, this comes at a cost: each reconstructed spectrum is necessarily within the convex hull of the measured spectra.
In particular, one can only reproduce luminance within this convex hull. 

\paragraph*{Applications} 
Metameric Blacks have been successfully used for camera calibration~\cite{alsam2007calibrating}, reflectance acquisition~\cite{morovic2006metamer,zhao2007image,lin2020physically}, or printing~\cite{morovic2011hans}.
However, this approach is too limiting for the artistic control of spectral assets, which is our main focus in this work.
Furthermore, working with discretized spectra has the additional drawback that spectral maxima are limited to occur at spectral bin locations, potentially preventing interesting effects.}

\subsection{Spectral Representations in Computer Graphics}

\paragraph*{Spectral upsampling.}
In computer graphics, the use of a spectral renderer requires to convert between colors and spectra~\cite{fascione2018manuka,weidlich2022practical}. Usually, the aim is to convert RGB textures (such as albedo maps, environment maps) to spectral textures with one spectral curve per texel -- a \textit{one-to-one mapping}~\cite{peercy1993linear,smits1999rgb,otsu2018reproducing,mallett2019spectral,jakob2019low,guarnera2022spectral,jendersie2021fast,todova2022wide}.
The difference between those approaches mostly resides in how spectra are built. 
For instance, \cite{smits1999rgb} and \cite{meng2015physically} optimize smooth spectra, \cite{peercy1993linear}and \cite{otsu2018reproducing} use a database to project colors, while \cite{jakob2019low} build a parametric family of spectra.
All these methods ensure that the resulting upsampled spectra are physically-plausible: they remain in the $[0,1]$ range along the spectral dimension.

\paragraph*{Spectral Compression.}
The storage of spectral curves raises additional difficulties. 
Parametric models (such as the one of \cite{jakob2019low}) limit storage requirements, even allowing for on-the-fly conversion of RGB assets.
However, they severely restrict the family of spectra that can be represented. 
An alternative is to decompose spectra using moments~\cite{peters2019using}. 
With this approach, it is possible to reconstruct a large family of spectra while keeping memory requirements in check.
The storage of spectra is orthogonal to our work, as we could choose to use any compression method to store the spectra produced by our approach.

\vspace{-5pt}
\paragraph*{Fluorescence.}
Spectra defined in the visible range can be extended to incorporate fluorescence effects~\cite{glassner1995model}. 
While it requires dedicated rendering algorithms~\cite{mojzik2018handling}, it expands the range of achievable appearances~\cite{jung2019wide}. While compact representations for fluorescent spectra have been recently introduced in computer graphics (e.g.,~\cite{Hua2021}), we restrict our approach to spectra in the visible range and put fluorescence aside.

\vspace{-5pt}
\subsection{Scope of this Work}

Our goal is to extend the computer graphics toolbox with a \textit{one-to-many} spectral upsampling method tailored to reflectance and transmittance.
Contrary to previous work in optics, we do not rely on convex combinations of measured spectra, since our focus is on the artistic control of color-changing effects, for vathochromism and metamerism alike.
As described in the next section, we overcome the difficulties raised by the null-space approach of Metameric Blacks by relying on a spectral Partition of Unity.

%% file: tex/sec_partition_of_unity.tex
\begin{figure}[b]
    \centering
    {\footnotesize
    \begin{tabular}{cc}
        \hspace{-15pt} \includegraphics[width=4.2cm]{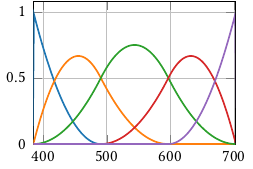} &
        \hspace{-5pt}  \includegraphics[width=4.2cm]{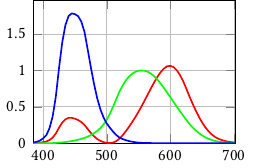} \\
        \textbf{a) Partition of Unity} &
        \textbf{b) CIE XYZ}
    \end{tabular}
    }
    \vspace{-5pt}
    \caption{
        Left: An example set of $K=5$ Partition of Unity (PU) basis functions of degree $2$ with regularly spaced knots. 
        Right: CIE sensitivity functions $\bar{x}(\lambda)$, $\bar{y}(\lambda)$ and $\bar{z}(\lambda)$ in red, green and blue.
        \label{fig:xyz_and_pu}
    }
\end{figure}

\section{Spectral Partition of Unity}
\label{sec:PU-geom}

In this section, we use a Partition of Unity to define a space of smooth spectra and show how these spectra are related to generalized barycentric coordinates in chromaticity space.

\vspace{-5pt}
\paragraph*{Partition of Unity.}
A Partition of Unity (PU) is a set of $K$ basis functions $B_k : U \rightarrow \mathbb{R}$ with $ k \in [0,K-1]$ such that:
\begin{align}
    \sum_k B_k(x) = 1, \forall x \in U.
\end{align}
We can use a weighted sum of these basis functions to reconstruct or approximate functions. 
A notable property of a PU is that bounded weights yield bounded reconstructed functions:
\begin{align}
    \forall k\in [0,K-1], \ w_k \in [0,1]  \Rightarrow f(x) = \sum_k w_k B_k(x) \in [0,1].
    \label{eqn:constraint_PU}
\end{align}

\paragraph*{Reconstructing Transmission Spectra.}
We use a PU created from non-uniform B-splines to produce reflectance or transmittance spectra.
The input domain is the set of visible wavelengths $U = [U_0, U_1] = [385\mbox{nm}, 700\mbox{nm}]$. 
The energy conservation constraint on reflectance and transmitance spectra is readily met through Equation~\ref{eqn:constraint_PU}. 
We will discuss the choice of the number $K$ of B-spline basis functions, their degree and the positions of their knots later in Section~\ref{sec:PU-design}.
In this section, for the purpose of illustration, we rely on $K=5$ bases of degree $2$ and uniformly spaced knots with knots at the boundaries of $U$ having a multiplicity of $3$, as shown in Figure~\ref{fig:xyz_and_pu}(left).
We also work with the sRGB color space.

\paragraph*{Geometric interpretation}
When intergrated with respect to the CIE sensitivity functions $\bar{x}(\lambda)$, $\bar{y}(\lambda)$ and $\bar{z}(\lambda)$ shown in Figure~\ref{fig:xyz_and_pu}(right), each basis function corresponds to a XYZ color:
\begin{align}
    \mathbf{B}_k 
    = \left[ \begin{array}{@{}c@{}} B_{k,X} \\ B_{k,Y} \\ B_{k,Z} \end{array} \right] 
    = \int B_k(\lambda) \mathbf{s}(\lambda) \mbox{d}\lambda,
\end{align}
with $\mathbf{s}(\lambda) = \left[ \bar{x}(\lambda), \bar{y}(\lambda), \bar{z}(\lambda) \right]^{\top}$.
Due to the linearity of reconstruction, a weighted sum of PU basis functions yields a XYZ color that is a  weighted sum of basis XYZ colors:
\begin{align}
    \mathbf{F} 
    = \left[ \begin{array}{@{}c@{}} F_X \\ F_Y \\ F_Z \end{array} \right] 
    = \int f(\lambda) \mbox{d}\lambda = \sum_k w_k \mathbf{B}_k.
    \label{eq:XYZ_coordinates_from_basis}
\end{align}

$\mathbf{F}$ may then be converted to the xyY color space.
Using Equation~\ref{eq:XYZ_coordinates_from_basis}, we directly obtain its luminance $F_Y = \sum_k w_k B_{k,Y}$.
Its chromaticity $\mathbf{c}$ is slightly more complicated. 
If we write $|F|=F_X+F_Y+F_Z$ and similarly $|B_k| = B_{k,X} + B_{k,Y} + B_{k,Z}$, it is given by:
\begin{equation*}
    \mathbf{c} = \dfrac{[F_X, F_Y]^{\top}}{|F|} = \dfrac{ \sum_k w_k \left[B_{k,X}, B_{k,Y}\right]^{\top} }{\sum_l w_l |B_l|}.
\end{equation*}
which may be rewritten as:
\begin{eqnarray}
    \mathbf{c} & = &  
    \sum_k a_k \mathbf{b}_k,\\ \label{eqn:chrom-linear}
    a_k & = & \frac{w_k |B_k|}{\sum_l w_l |B_l|}. \label{eqn:chrom-bary-coords}
    \label{eq:chroma_from_basis}
\end{eqnarray}
where the $\mathbf{b}_k=\tfrac{[B_{k,X}, B_{k,Y}]^{\top}}{|B_k|}$ denote basis chromaticities. \\

\begin{figure}[t]
    \centering
    \vspace{-5pt}
    \begin{tikzpicture}[font=\small]
        \node at (0.0, 0.0) {\includegraphics[width=\linewidth]{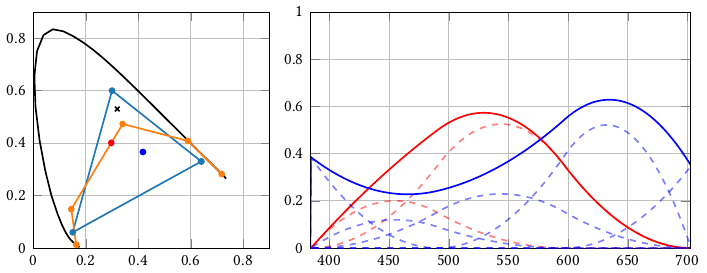}};
        \node at (-2.5, -1.9) {\change{}{\textbf{a) Chromaticity space}}};
        \node at ( 1.8, -1.9) {\change{}{\textbf{b) Reconstructed spectra from PU}}};
    \end{tikzpicture}
    \vspace{-17pt}
    \caption{
        A PU basis $B_k(\lambda)$ defines a gamut in chromaticity space ((a) orange polygon) where each vertex $\mathbf{b}_k$ is a basis element. 
        Any point outside that gamut (e.g., the black cross) is not achievable with the chosen basis, even though it might be inside the sRGB gamut (blue triangle).
        A spectrum defined by two non-null contiguous basis coefficients (e.g., plain red curve in (b) with individual PU contributions in dashed) yields a chromaticity on the basis gamut boundary (red point).
        More general spectra (e.g., blue curve in (b)) yield a chromaticity inside the basis gamut (blue point in (a)).
        \label{fig:chromaticity_and_pu}
        \vspace{-15pt}
    }
\end{figure}

Our key observation is thus that the chromaticity $\mathbf{c}$ of a spectrum given by a vector $\mathbf{w}$ of basis coefficients is obtained as a linear combination of basis chromaticities $\mathbf{b}_k$ where the weights $a_k$ correspond to \emph{homogeneous barycentric coordinates}.
This is illustrated in Figure~\ref{fig:chromaticity_and_pu}, where the basis chromaticities $\mathbf{b}_k$ form a gamut of colors achievable through a given choice of basis functions $B_k(\lambda)$.
Depending on that choice, the gamut may only partially overlap the sRGB gamut: this means that there is no $\mathbf{w}$ that can achieve a chromaticity outside the basis gamut.
A vector $\mathbf{w}$ with only two non-zero contiguous coefficients yields a unique chromaticity point on the gamut boundary, since then only a contiguous pair of barycentric coordinates is non-zero.
However, in all other cases, there will be multiple coefficient vectors $\mathbf{w}$ that map to the same chromaticity point $\mathbf{c}$.
This is because for $K>3$, the set of $a_k$ describes \emph{generalized} barycentric coordinates of $\mathbf{c}$, and is thus \emph{not} unique.
In the next section, we show how to invert this many-to-one mapping.

%% file: tex/sec_nullspace_sampling.tex
\section{One-to-many mapping}
\label{sec:PU-explore}

Our goal in this section is to find the equivalence class of basis coefficients $\mathbf{w}$ that yields a target chromaticity $\mathbf{c}$ and luminance $F_Y$.
We do this in two stages: we first find the set of generalized barycentric coordinates that achieves the target chromaticity $\mathbf{c}$; then we show how this maps to an equivalence class of basis coefficients, a subset of which achieves the target luminance $F_Y$.

\vspace{-10pt}
\subsection{Achieving chromaticity} 
A first condition is that $\mathbf{c}$ must lie inside the basis gamut or on its boundary.
The target chromaticity may then be expressed in terms of generalized homogeneous barycentric coordinates, using:
\begin{equation}
\label{eqn:general_barycentric}
    \left[\begin{array}{@{}cccc@{}}
        1 & 1 & \cdots & 1 \\
        b_{0,x} & b_{1,x} & \cdots & b_{K-1,x} \\
        b_{0,y} & b_{1,y} & \cdots & b_{K-1,y}
    \end{array}\right]
    \left[\begin{array}{@{}c@{}}
        a_0 \\ a_1 \\ \cdots \\ a_{K-1}
    \end{array}\right] =
    \left[\begin{array} {@{}c@{}}
        1 \\ c_x \\ c_y
    \end{array}\right],
\end{equation}
with $\mathbf{b}_k = [b_{k,x},b_{k,y}]^{\top}$ and $\mathbf{c} = [c_x,c_y]^{\top}$.

Since $\mathbf{c}$ is in the basis gamut, there is at least one triplet of bases whose chromaticity coordinates define a triangle that contains $\mathbf{c}$. 
Let's assume that these basis are the first three (one can always re-order the bases to yield such a  configuration).
One solution to Equation~\ref{eqn:general_barycentric} is then $[a_0, a_1, a_2, 0, \cdots, 0]^\top = [\mathbf{a}_T^{\top}, \mathbf{0}]^\top$, with $\mathbf{a}_T$ the vector of triangular barycentric coordinates.
Other solutions may then be obtained by adding perturbations to that vector, which may be written $[a_0-\Delta a_0,a_1-\Delta a_1,a_2-\Delta a_2,a_3,\cdots,a_{K-1}]^\top = [(\mathbf{a}_T-\Delta \mathbf{a})^{\top}, \mathbf{a}_F^ \top]^\top$, where $\mathbf{a}_F$ is a (K-3)D vector of barycentric coordinates that represent degrees of freedom to navigate the space of solutions, and $\Delta \mathbf{a}$ is a 3D offset vector used to preserve the homogeneous barycentric coordinate constraint.

\begin{figure*}[!h]
    \centering
    \begin{tikzpicture}[font=\small]
        \node{\includegraphics[width=\linewidth]{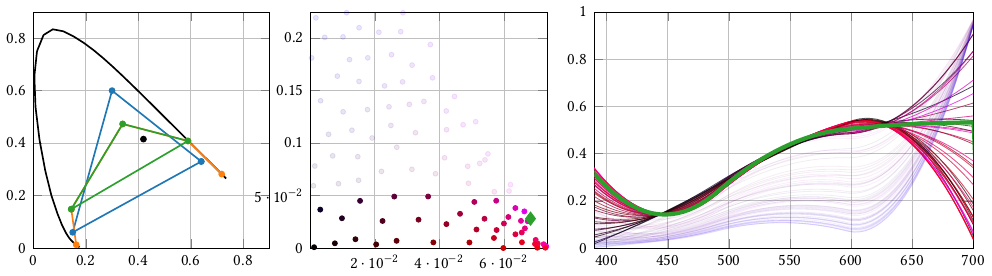}};
        \node at (-6.3, -2.65) {\change{}{\textbf{a) Barycentric coordinates}}};
        \node at (-6.3, -2.95) {\change{}{\textbf{  in chromaticity space }}};
        \node at (-1.3, -2.65) {\change{}{\textbf{b) Sampling barycentric}}};
        \node at (-1.3, -2.95) {\change{}{\textbf{ degrees-of-freedom}}};
        \node at ( 5.4, -2.65) {\change{}{\textbf{c) Generated spectra}}};
    \end{tikzpicture}
    \vspace{-15pt}
    \caption{\textbf{One-to-many mapping.}
    The target color is given by its chromaticity $\mathbf{c}=[0.41, 0.42]^{\top}$ and luminance $F_Y=0.57$.
    Left: a triangle (in green) enclosing the target chromaticity $\mathbf{c}$ (black dot) is picked  among the basis gamut (in orange).
    Middle: the remaining two basis constitute degrees of freedom, which are randomly sampled using our iterative procedure based on barycentric coordinates. 
    Observe the presence of boundaries in this barycentric space, which are required to keep basis coefficients in the $[0,1]$ range.
    Right: the equivalence class of spectra achieving $\mathbf{c}$ are retrieved from the degrees of freedom (we use matching colors).
    Transparent barycentric samples (middle) and spectra (right) indicate that the target luminance $F_Y$ is not achieved.
    The solution of maximum luminance in the equivalence class is shown with a green diamond in barycentric space (middle), and its spectrum is drawn in green (right).
        \label{fig:explore_bary}
    }
\end{figure*}

Let us now rewrite Equation~\ref{eqn:general_barycentric} with the following matrix form: $\left[ T \ F \right] \mathbf{a} = [1, \mathbf{c}^{\top}]^{\top}$, where $T$ (resp. $F$) is the matrix corresponding to the first $3$ (resp. last $K-3$) columns of the left hand side matrix, and $\mathbf{a}$ is the vector of generalized barycentric coordinates.
Since we also have $T \mathbf{a}_T = [1, \mathbf{c}^{\top}]^{\top}$, it follows that:
\begin{equation}
    \label{eqn:general_barycentric2}
    T \Delta \mathbf{a} = F \mathbf{a}_F.
\end{equation}
Now in order to navigate the space of solutions, we need bounds on $\mathbf{a}_F$. 
Because all its coefficents are barycentric coordinates, we already know that $\mathbf{0} \le \mathbf{a}_F \le \mathbf{1}$, with the lower bound trivially corresponding to a zero offset vector (see Equation~\ref{eqn:general_barycentric2}).
The upper bound is not a sufficient condition since we must also make sure that $\mathbf{0} \le \mathbf{a}_T - \Delta \mathbf{a} \le \mathbf{1}$, or in terms of the offset vector: $\mathbf{a}_T-\mathbf{1} \le \Delta \mathbf{a} \le \mathbf{a}_T$. 
Using Equation~\ref{eqn:general_barycentric2} yields the following vector inequality:
\begin{equation}
\label{eqn:general_barycentric_bounds}
    \mathbf{a}_T - \mathbf{1}\le M \mathbf{a}_F \le \mathbf{a}_T,
\end{equation}
where $M=T^{ -1} F$ is a $3 \times (K-4)$ matrix.
Note that since $M$ may contain negative coefficients, the lower bound in Equation~\ref{eqn:general_barycentric_bounds} may end up being used to define the upper bound on $\mathbf{a}_F$.

We rely on an iterative approach to characterize the whole set of solutions by considering each coefficient of $\mathbf{a}_F$ in turn.
Let us start with $a_3$, and assume that $a_{4..K-1} = 0$.
Equation~\ref{eqn:general_barycentric_bounds} now becomes $ \mathbf{a}_T - \mathbf{1} \le \mathbf{m}_0 a_3 \le \mathbf{a}_T$, with $\mathbf{m}_0=[m_{00}, m_{10}, m_{20}]^\top$ the first column of $M$.
The constraints on offsets are then met by navigating $a_3$ in the $[0,a_3^{\max}]$ interval, with the upper bound given by:
\begin{equation}
\label{eqn:general_barycentric_bounds_3}
a_3^{\max}=\min_{i \in \{0,1,2\}} \frac{a_i+H(m_{i0})-1}{m_{i0}}.
\end{equation}
The Heaviside function $H(m)$ is used to take the sign of each matrix component into account: when $m\le0$ (resp. $m>0$), the lower (resp. upper) bound is considered.
Having chosen the $n-1$ first coefficients of $\mathbf{a}_F$, the upper bound for the $n$th coefficient -- assuming the remaining ones are zero -- is obtained with a similar formula:
\begin{equation}
\label{eqn:general_barycentric_bounds_k}
a_{3+n}^{\max}=\min_{i \in \{0,1,2\}} \frac{a_i+H(m_{in})-1-\sum_{l=0}^{n-1} m_{il} \, a_{3+l}}{m_{in}}.
\end{equation}

In the general case, a valid solution  $\forall n \in [0..K-4]$ must ensure:
\begin{equation}
\label{eqn:general_barycentric_bounds_all}
a_{3+n} \le \min_{i \in \{0,1,2\}} \frac{a_i+H(m_{in})-1-\sum_{l \neq n} m_{il} \, a_{3+l}}{m_{in}}.
\end{equation}
For each vector $\mathbf{a}_F$, the offset vector is computed by $\Delta \mathbf{a} = M \mathbf{a}_F$, which yields a generalized homogeneous coordinates vector $\mathbf{a}$ that achieves the target chromaticity $\mathbf{c}$.
Figure~\ref{fig:explore_bary} illustrates that process. 
A triangle that encloses $\mathbf{c}$ is first selected; then the space of degrees of freedom $\mathbf{a}_F$ is sampled randomly to yield a family of spectra.

\subsection{Achieving luminance}
Given a vector of generalized barycentric coordinates $\mathbf{a}$, we now need to invert Equation~\ref{eqn:chrom-bary-coords} to retrieve basis coefficients $\mathbf{w}$.
Since any pair of basis coordinates $(a_i,a_j)$ is related by an equation of the form $a_i w_j |B_j| = a_j w_i |B_i|$, the corresponding basis coefficients span a $K$D line.
If we pick an arbitrary non-zero barycentric coordinate -- say $a_0$ -- then $\mathbf{w}$ may be expressed as a function of $w_0$:
\begin{equation}
\label{eqn:inverse-poly-line}
    \mathbf{w}(w_0) =
    \left[\begin{array}{@{}c@{}} 1 \\ \frac{a_1 |B_0|}{a_0 |B_1|} \\ \cdots \\ \frac{a_{K-1} |B_0|}{a_0 |B_{K-1}|} \end{array} \right] w_0 = L w_0, \ \ \ w_0 \in \left(0, w_0^{\max}\right],
\end{equation}
where the upper bound $w_0^{\max} = \min \left\{1,\frac{a_0 |B_1|}{a_1 |B_0|},\cdots,\frac{a_0 |B_{K-1}|}{a_{K-1} |B_0|}\right\}$ is set to ensure that $\mathbf{0} \le \mathbf{w} \le \mathbf{1}$.

The KD line of solutions may now be restricted to a single solution by the target luminance constraint, which we write $\mathbf{w}(w_0)^{\top} \mathbf{B}_y = F_Y$, with $\mathbf{B}_y=[B_{0,Y}, \cdots, B_{K-1,Y}]^{\top}$ the vector of Y coefficients of basis colors.

Using Equation~\ref{eqn:inverse-poly-line}, the value of $w_0$ that \emph{potentially} achieves the target luminance $F_Y$ is:
\begin{equation}
\label{eqn:inverse-luminance}
    w_0^{\star} = \frac{F_Y}{L^\top \mathbf{B}_y}.
\end{equation}
$F_Y$ is effectively achieved if and only if $w_0^{\star} \le w_0^{\max}$.
For that reason only a (possibly empty) subset of barycentric coordinates $\mathbf{a}$ permits to achieve the target luminance.
This is shown in Figure~\ref{fig:explore_bary}: only fully-opaque barycentric samples and their associated spectra achieve $F_Y$ in practice.

However, this subset may be enlarged.
Indeed, relying on the bounded property of PU (Equation~\ref{eqn:constraint_PU}) remains conservative: in some instances, we may use basis coefficients greater than $1$ and still obtain energy-conserving spectra. 
This means that $\mathbf{w}$ may be scaled in post process to increase the luminance of the reconstructed spectrum.
Assuming that $F_Y$ is not achieved (i.e., $\mathbf{w}(w_0^{\max})^{\top}\mathbf{B}_Y < F_Y$), we may thus obtain a closer solution in terms of luminance by using:
\begin{equation}
    \mathcal{W}(\mathbf{w}) = \frac{\mathbf{w}(w_0^{\max})}{\max \left(f^{\max}, \tfrac{\mathbf{w}(w_0^{\max})^{\top}\mathbf{B}_Y}{F_Y}\right)},
\end{equation}
where $f^{\max} = \max_{\lambda} f(\lambda)$, and $1/f^{\max}$ represents the margin by which the spectrum $f(\lambda)$ is allowed to be scaled.

Finally, it would be useful to know \emph{a priori} whether there exists at least one vector $\mathbf{w}$ of basis coefficients that achieves both the target chromaticity and luminance.
A \emph{conservative} solution is to rely on the vector $\mathbf{\overline{w}}$ that maximizes luminance under the constraint given by Equation~\ref{eqn:general_barycentric}, then to check whether $\mathcal{W}(\mathbf{\overline{w}})^{\top} \mathbf{B}_y \ge F_Y$.
The vector $\mathbf{\overline{w}}$ is found by solving the following linear programming problem:  
\begin{eqnarray}
    \label{eqn:maxLum-weight}
    \mathbf{\overline{w}} & = & \max_{\mathbf{w}} \ \mathbf{w}^{\top} \mathbf{B}_y, \\ \label{eqn:maxLum-wConstraint}
    \mbox{subject to} && \mathbf{0} \le \mathbf{\overline{w}} \le \mathbf{1} \\ 
    \mbox{and } && A \ \mathbf{\overline{w}} = \mathbf{0},
\end{eqnarray}
where $A$ is obtained by rewriting Equation~\ref{eqn:general_barycentric} in terms of $\mathbf{w}$:
\begin{equation}
    A = [T \ F] \ \mbox{diag}(\mathbf{|B|})^{\top} - \mathbf{|B|}^{\top} \left[\begin{array}{@{}c@{}} 1 \\ \mathbf{c} \end{array}\right],
\end{equation}
 where we have used $\mathbf{a}=\tfrac{\mbox{diag}(\mathbf{|B|})^{\top}\mathbf{w} }{ \mathbf{|B|}^{\top}\mathbf{w}}$ and $\mathbf{|B|} = [|B_0|, \cdots, |B_{K-1}|]^{\top}$.
 An exact solution could be obtained by replacing $\mathbf{\overline{w}}$ by $\mathcal{W}(\mathbf{\overline{w}})$ in Equation~\ref{eqn:maxLum-wConstraint}, at the cost of a more expensive computation time.

 Once we have found $\mathbf{\overline{w}}$ (green spectrum in Figure~\ref{fig:explore_bary}), it is trivial to retrieve the corresponding barycentric coordinates $\mathbf{\overline{a}}$ and degrees of freedom $\mathbf{\overline{a}}_F$ (green diamond in Figure~\ref{fig:explore_bary}).
 We use $\mathbf{\overline{w}}$ by default during editing (see the supplemental video).

%% file: tex/sec_basis_design.tex
\section{Basis design}
\label{sec:PU-design}

\begin{figure}[t]
    \centering
    \begin{tikzpicture}[font=\footnotesize] 
        \node[inner sep=0pt, ] (nb5) { \includegraphics[width=0.245\linewidth]{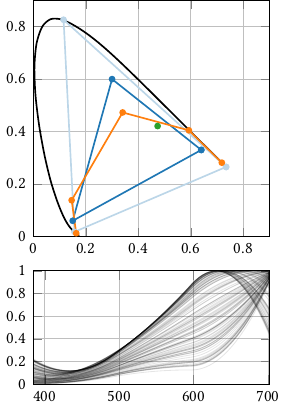} };
        \node[above=0pt of nb5] {$K = 5$};
        \node[inner sep=0pt, right=0pt of nb5] (nb7) { \includegraphics[width=0.245\linewidth]{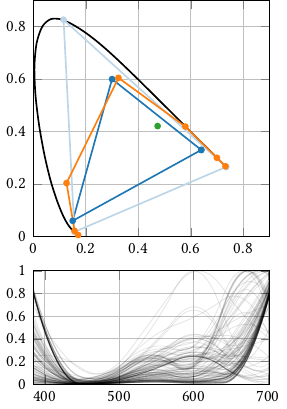} };
        \node[above=0pt of nb7] {$K = 7$};
        \node[inner sep=0pt, right=0pt of nb7] (nb9) { \includegraphics[width=0.245\linewidth]{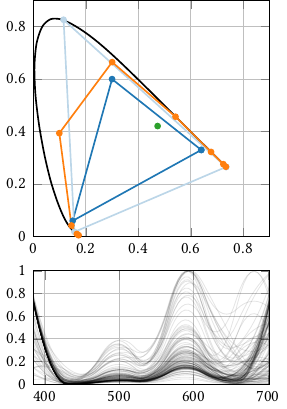} };
        \node[above=0pt of nb9] {$K = 9$};
        \node[inner sep=0pt, right=0pt of nb9] (nb11) { \includegraphics[width=0.245\linewidth]{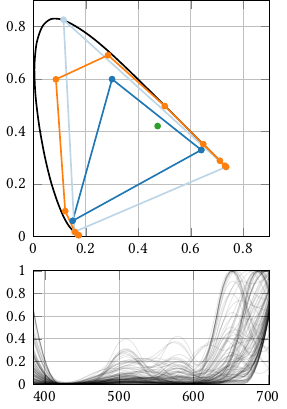} };
        \node[above=0pt of nb11] {$K = 11$};
    \end{tikzpicture}
    \caption{
        \textbf{Basis gamut w.r.t. $K$.} 
        For the same chromaticity constraint (green dot, top row), we display randomly generated spectra (bottom row) when increasing the numbers of basis functions (from left to right $K = \{5,7,9,11\}$). {}{We compare the basis gamut to both sRGB (blue) and Adobe Wide Gamut RGB (light blue).}
        \label{fig:basis_design_nb_basis}
    }
\end{figure}

Until now, we have relied on a small number of basis functions ($K=5$) for illustration purposes.
Increasing the number $K$ of bases has the effect of increasing the size of the equivalence class.
As shown in Figure~\ref{fig:basis_design_nb_basis}, 
this is due to the basis gamut, which encompasses a larger area of the chromaticity diagrams when $K$ is increased.
This has two effects on the \emph{expressivity} of a given basis.
First, any chromaticity in a given RGB gamut {}{(we consider sRGB and Adobe Wide Gamut RGB in the following)} can be achieved when it is encompassed by the basis gamut. 
{}{Increasing the number of bases extends the latter as shown in Figure~\ref{fig:basis_design_nb_basis} (top row)}. 
Second, chromaticities on the gamut boundary map to a single pair of non-zero barycentric coordinates; hence a basis gamut larger than the chosen RGB gamut ensures to avoid these singular equivalence classes.

However, increasing the number  $K$ of bases cannot be done without limits, since plausible reflectance and transmittance spectra should be \emph{smooth}.
In addition, a smaller number of bases might be desirable for memory considerations.\\

\vspace{-7pt}
In this section, from the geometric interpretation of previous sections, we design a set of PU basis functions that finds a tradeoff between expressivity and smoothness constraints.
We keep degree of $2$ throughout as there is no need to ensure $C^2$ (or higher) continuity, and low degree splines prevent us from over-fitting issues.

\vspace{-5pt}
\paragraph*{Knots warping}
Besides increasing $K$, we may also control the position of basis knots.
To this end, we use a two-parameter family of warping functions to alter the uniform distribution of knots along the $U = [U_0, U_1]$ interval.
We use a warping function $C_{s,p}: [0,1] \to [0,1]$ introduced by \cite{hise2020}: 
\begin{equation}
    C_{s,p}(x) = \left\{
    \begin{array}{@{}l@{\hspace{5pt}}r@{}}
        \frac{x^{c}}{p^{c-1}} & \mbox{~ if~ } x \in [0,p],\\
        1 - \frac{(1-x)^{c}}{(1-p)^{c-1}} & \mbox{~ otherwise},
    \end{array}
    \right.
\end{equation}
with $c = \tfrac{2}{1+s}-1$.
The two parameters $(s,p) \in [0,1]^2$ control the strength of warping and the position where most of the warping occurs.
A sequence of warped knots $\{\kappa_k\}$  is then produced using $\kappa_k = U_0 + C_{s,t}(u_k) (U_1 - U_0)$, where the $u_k$ form a uniform sequence of values in the $[0,1]$ range.
Even though a set of $K$ B-spline basis functions of order $2$ requires $K+3$ knots, we ignore the first and last two since the boundary knots have a multiplicity of $3$.
Hence we obtain $K-1$ knots, with $\kappa_0=U_0$ and $\kappa_{K-1}=U_1$ as desired. 

\begin{figure}[t]
    \centering
    \begin{tikzpicture}[font=\footnotesize]
        \node[inner sep=0pt] (A) { \includegraphics[width=\linewidth]{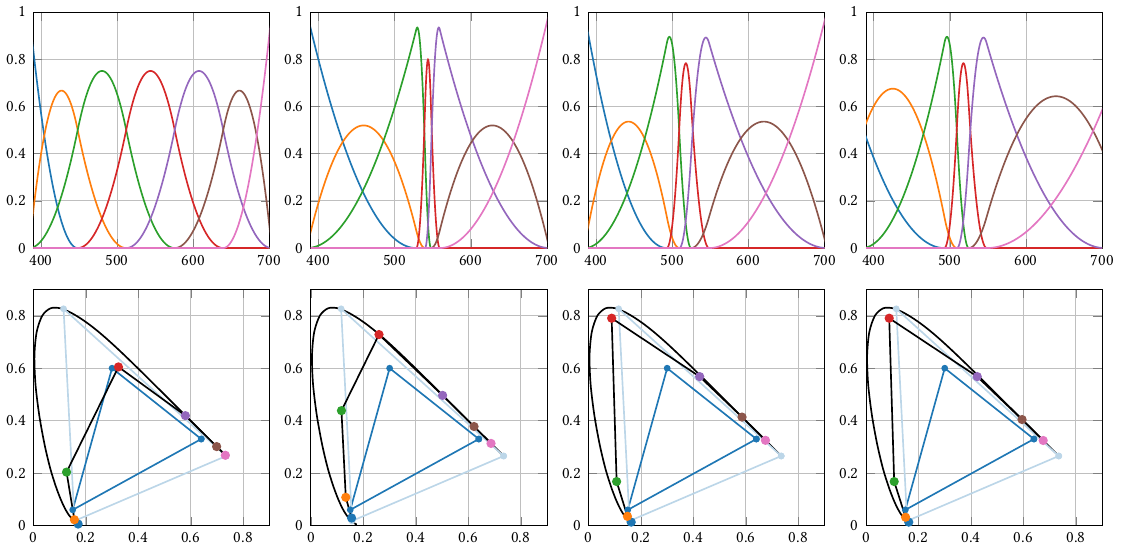}};
        \node[anchor=center] (t0) at (-3.1,2.2) { \textbf{ {no warping} }};
        \node[anchor=center] (t0) at (-1.0,2.5) { \textbf{ {$s = 0.8$} }};
        \node[anchor=center] (t0) at (-1.0,2.2) { \textbf{ {$p = 0.5$} }};
        \node[anchor=center] (t0) at ( 1.0,2.5) { \textbf{ {$s = 0.66$} }};
        \node[anchor=center] (t0) at ( 1.0,2.2) { \textbf{ {$p= 0.39$} }};
        \node[anchor=center] (t0) at ( 3.1,2.2) { \textbf{ {+boundary knots} }};
    \end{tikzpicture}
    \caption{
        \textbf{Knots warping.}
        Warping basis knots alters the basis gamut, here for a set of $K=7$ basis functions.
        We assign to each basis a color to clearly locate it in chromaticity space.
        A strong warping (second column) tends to widen the gamut considerably, but results in one very narrow basis function. 
        Adjusting the position parameter $p$ (third column) achieves an even wider gamut with a smaller strength parameter $s$, which results in less narrow bases and permits to capture most of the XYZ space.
        Displacing the boundary knots outside of the spectral interval $U$ (last column) has a negligible effect on the basis gamut, even though pairs of basis functions at boundaries are significantly modified.
        \label{fig:knots_warping}
    }
\end{figure}

Figure~\ref{fig:knots_warping} shows the effect of knots warping on $K=7$ basis functions and the corresponding basis gamut, demonstrating that knots warping helps achieve a wider gamut without having to increase the number $K$ of basis functions.
The figure also shows that modifying the first and last two knots to be outside of the $U$ interval has a negligible effect on the basis gamut, which is due to the small values of color matching functions around these boundaries (see Figure~\ref{fig:xyz_and_pu}(left)).
We choose to offset these boundary knots by $100$nm outside of $U$, which produces more physically-plausible results when observed outside of the visible range since the reconstructed spectra then gently fade to zero outside of $U$.

\paragraph*{Expressivity-smoothness trade-offs}
We need to devise metrics to quantify the degree of expressivity of a set of basis functions, as well as the smoothess of the spectra it is able to produce, depending on the positions of its knots and the number $K$ of bases.

\begin{figure}[t]
    \begin{tikzpicture}[font=\footnotesize]
        \node[inner sep=0pt] (A) {\includegraphics[width=\linewidth]{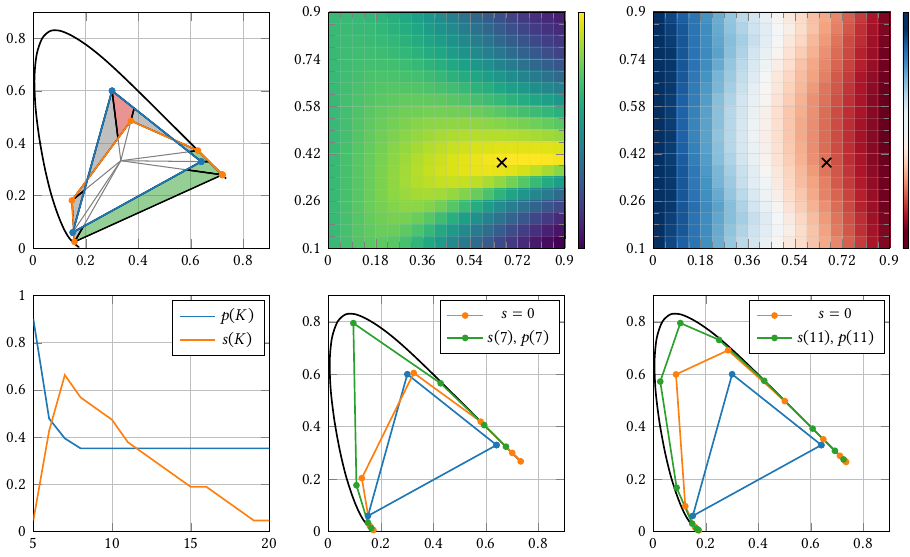}};
        \node (tA) at (-0.1,2.7) { \textbf{{excess area}} };
        \node (tB) at (3,2.725) { \textbf{{smoothness}} };
    \end{tikzpicture}
    \caption{
        \textbf{Warping optimisation.}
        Top row: the excess area $\mathcal{A}$ is computed by tesselating the region between the RGB and basis gamuts into quads (left) and adding their signed areas (positive in green, negative in red, mixed in gray).
        We computed $\mathcal{A}$ for several values of the $(s,p)$ warping parameters (middle) as well as a smoothness criterion $\mathcal{S}$ (right).
        Bottom row: our $(s(K),p(K))$ trade-offs for various numbers $K$ of basis functions, and two example basis gamuts before (orange) and after (green) warping.
        We plot $(s(7),p(7))=(0.66,0.39)$ with black crosses in criteria maps.
        \label{fig:knots_optim}
        \vspace{-20pt}
    }
\end{figure}

An expressive basis requires a wide gamut that encompasses the {}{chosen RGB gamut as much as possible.} 
We thus rely on what we call the excess area $\mathcal{A}$, which is the signed area between the basis and sRGB gamuts, normalized by the area between the horseshoe-shaped chromaticity gamut and the RGB gamut.
This area is computed by tesselating the region between the RGB and basis gamuts into quads (see Figure~\ref{fig:knots_optim}(left)).

Basis smoothness is directly related to the smoothness of individual basis functions, which depends on both the number $K$ of bases and their knots $\{\kappa_k\}$.
We compute the smoothness of a basis set as $\mathcal{S} = \min_k \mbox{FWHM}_k$, where $\mbox{FWHM}_k$ is the full width at half maximum of the $k$th basis. 

As shown in Figure~\ref{fig:knots_optim}, for $K=7$ bases {}{and a sRGB spectrum}, the excess area $\mathcal{A}$ and smoothness $\mathcal{S}$ criteria evolve differently as a function of $(s,p)$, the parameters of the knots warping function.
How this pair of criteria is balanced is arbitrary.
In this paper, we usually first pick a number $K$ of basis functions, and then brute-force find the warping parameters that maximize $\mathcal{A}$ under the constraint that $\mathcal{S} < 20$nm.
We indicate such a $(s,p)$ pair for $K=7$ in Figure~\ref{fig:knots_optim} by a black cross.

%% file: tex/sec_applications.tex
\section{Applications}
\label{sec:apps}

We now present application scenarios where we use our one-to-many mapping to find spectra with interesting visual appearance.
Unless otherwise specified, we use warped basis functions with $(s,p)$ parameters determined as described in the previous section.

\subsection{Reproducing Vathochromism}
\label{sec:apps-vatho}

Figure~\ref{fig:teaser} demonstrates a reproduction of the Usambara effect using our approach. 
We use $K=11$ warped basis functions and sample the equivalence class of spectra that achieves the target chromaticity $\mathbf{c}=[0.38, 0.45]^{\top}$ and luminance $F_y=0.46$.
All such spectra are considered as transmittance spectra at a unit optical depth, which is related to the extinction coefficient $\sigma_t$ of a medium by $T_1(\lambda) = e^{-\sigma_t(\lambda)}$.
The Beer-Lambert-Bouguer law at increasing depths $d$ is then given by $T_d(\lambda) = T_1(\lambda)^d$.
For each sample of the equivalence class, we then integrate the corresponding $T_d$ over color matching functions and plot the resulting transmittance curve in the chromaticity diagram.
A pair of examples is shown in Figure~\ref{fig:teaser}(c), where we have picked two instances of the class that reproduce the Usambara effect -- here with an orange color at large optical depths.
For rendering, we need to specify $\sigma_a$ and $\sigma_s$, the absorption and scattering coefficients.
In Figure~\ref{fig:teaser}(b), we use $\sigma_s(\lambda)=T_1(\lambda)$ to achieve the target color on single scattering, which yields $\sigma_a(\lambda) = -\log T_1(\lambda) - T_1(\lambda)$.

\begin{figure*}[t]
    \begin{tikzpicture}[font=\footnotesize]
        \node[inner sep=0pt] (chroma) { \includegraphics[height=4.3cm]{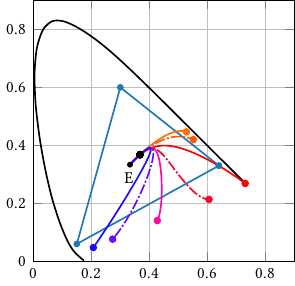} };
        \node[inner sep=0pt,right=5pt of chroma.north east, anchor=north west] (r0) {\includegraphics[width=1.7cm]{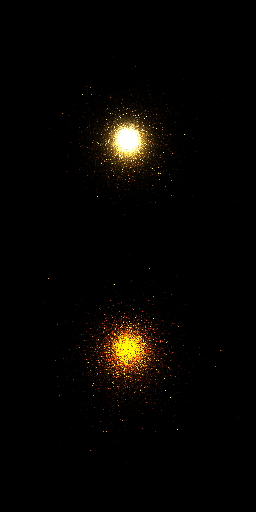}};
        \node[inner sep=0pt,right=2pt of r0] (r1) {\includegraphics[width=1.7cm, frame]{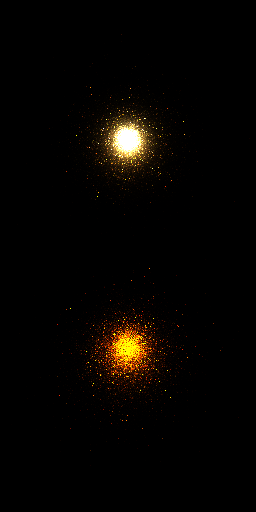}};
        \node[inner sep=0pt,right=2pt of r1] (r2) {\includegraphics[width=1.7cm, frame]{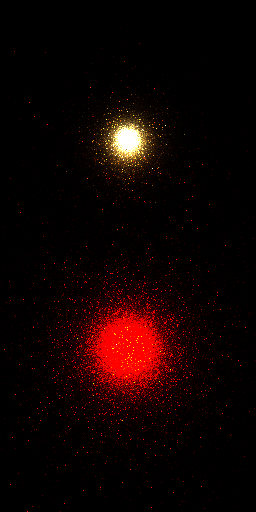}};
        \node[inner sep=0pt,right=2pt of r2] (r3) {\includegraphics[width=1.7cm, frame]{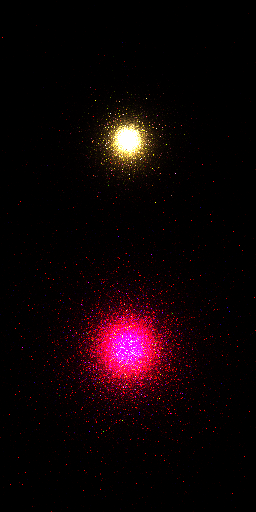}};
        \node[inner sep=0pt,right=2pt of r3] (r4) {\includegraphics[width=1.7cm, frame]{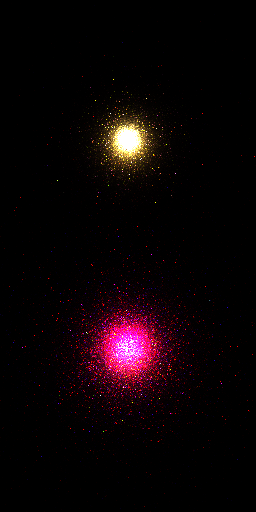}};
        \node[inner sep=0pt,right=2pt of r4] (r5) {\includegraphics[width=1.7cm, frame]{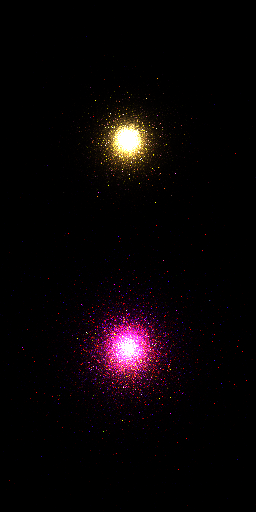}};
        \node[inner sep=0pt,right=2pt of r5] (r6) {\includegraphics[width=1.7cm, frame]{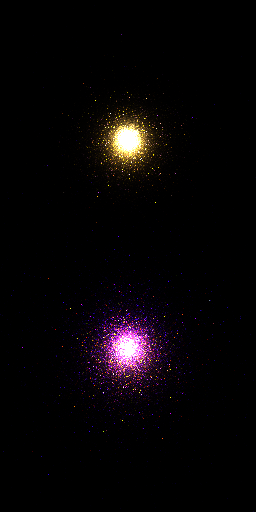}};

        \node[inner sep=0pt,below=1pt of r0] (s0) { \includegraphics[width=1.7cm, height=0.5cm]{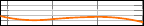} };
        \node[inner sep=0pt,below=1pt of r1] (s1) { \includegraphics[width=1.7cm, height=0.5cm]{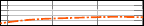} };
        \node[inner sep=0pt,below=1pt of r2] (s2) { \includegraphics[width=1.7cm, height=0.5cm]{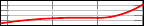} };
        \node[inner sep=0pt,below=1pt of r3] (s3) { \includegraphics[width=1.7cm, height=0.5cm]{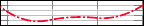} };
        \node[inner sep=0pt,below=1pt of r4] (s4) { \includegraphics[width=1.7cm, height=0.5cm]{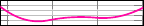} };
        \node[inner sep=0pt,below=1pt of r5] (s5) { \includegraphics[width=1.7cm, height=0.5cm]{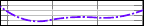} };
        \node[inner sep=0pt,below=1pt of r6] (s6) { \includegraphics[width=1.7cm, height=0.5cm]{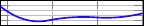} };
    \end{tikzpicture}
    \vspace{-10pt}
    \caption{
    \textbf{Parametrization and unit tests.} 
    We use our parametrization based on representative spectra to navigate through equivalence classes of transmittance spectra.
    The resulting spectra (colored by their hue at a large depth) are used for both absorption and scattering coefficients for a slab of homogeneous medium. 
    The slab is lit by a point light source in front, and another from behind to distinguish first-order scattering (top row) from high-order scattering (bottom row). 
    We scaled the medium to highlight $d=10$ optical depth points (colored dots in the chromaticity diagram).
    \label{fig:results_usambara_unit_test}
    \vspace{-15pt}
    }
\end{figure*}

\change{}{\paragraph*{Parameterizing the equivalence class}}
In the previous example, randomly sampling the equivalence class and then picking spectra that achieve the desired effects only provides  indirect control over achieved colors at a optical depths $d>1$.
For some applications, a more direct control might be desired. 
Unfortunately, depending on the choice of basis, not all color appearance choices can be achieved.
We provide a more direct control by parametrizing the \change{}{equivalence class} for the specific case of vathochromic transmittance spectra, which we illustrate on a series of unit tests in Figure~\ref{fig:results_usambara_unit_test}.
The main idea is to pick an \emph{a priori} set of representative spectra from the equivalence class, order them in chromaticity space, and interpolate them to navigate through a subset of relevant spectra.
We use spectra formed by all triangles that contain the target chromaticity $\mathbf{c}$, which naturally tend to result in distinct color appearance since they only rely on three basis functions.
We then sample their transmittance curves at an arbitrary optical depth, and order the representative spectra in clockwise order around the equiluminant point $E=(\tfrac{1}{3},\tfrac{1}{3})$ according to chromaticity samples, as illustrated in Figure~\ref{fig:results_usambara_unit_test}(left).
As demonstrated in the accompanying video, when the user specifies a hue, we interpolate among the two closest transmittance curves in the chromaticity diagram.

\paragraph*{Unit tests}
The remainder of Figure~\ref{fig:results_usambara_unit_test} shows a test scene composed of a slab of homogeneous transparent and scattering medium, lit by two white point light sources: one behind and one in front\footnote{Such scenes are typically long to converge in a spectral path tracer. We will provide more converged results in a final version of the paper.}.
As in Figure~\ref{fig:teaser}, the absorption and scattering coefficients are determined from $T_1(\lambda)$ for each of the seven representative spectra of the equivalence class to render test images.
The optical depth of light paths coming from behind is typically short, and exhibits the target chromaticity $\mathbf{c}$ for all tests as expected.
Light paths that come from the light in front instead need to be scattered to reemerge toward the camera and exhibit target hues at greater optical depths.

\begin{figure}[h]
    \centering
    \begin{tikzpicture}[font=\tiny]
        \node[inner sep=0pt]                   (A00) { \includegraphics[width=0.3\linewidth, frame]{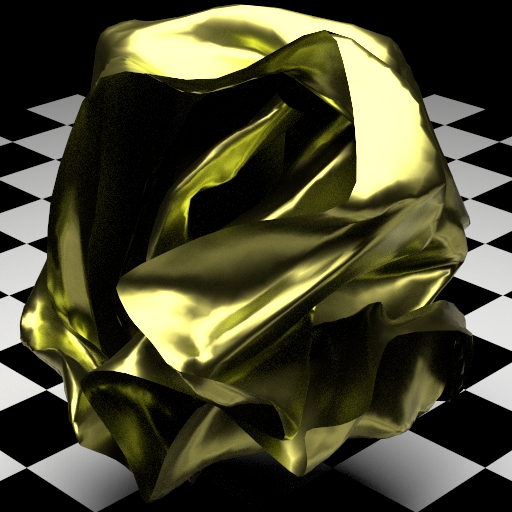} };
        \node[inner sep=0pt, right=1pt of A00] (A02) { \includegraphics[width=0.3\linewidth, frame]{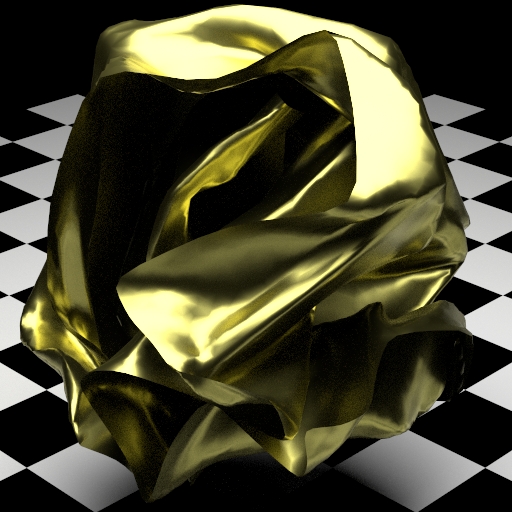} };
        \node[inner sep=0pt, right=1pt of A02] (A03) { \includegraphics[width=0.3\linewidth, frame]{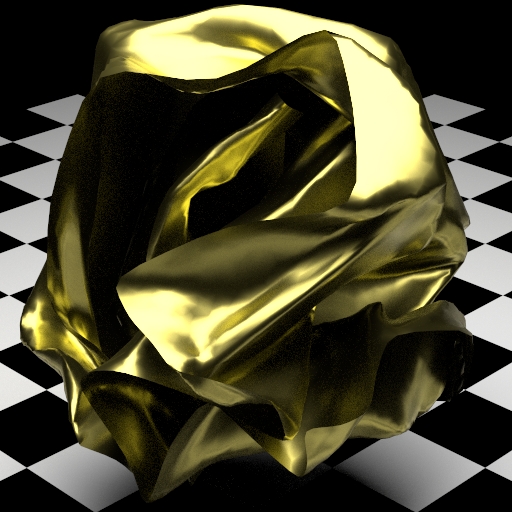} };
        \node[rotate=90, left=0pt of A00, anchor=south] (Ds) { {\textbf{Single Scattering}} };

        \node[inner sep=0pt, below=2pt of A00] (A10) { \includegraphics[width=0.3\linewidth, frame]{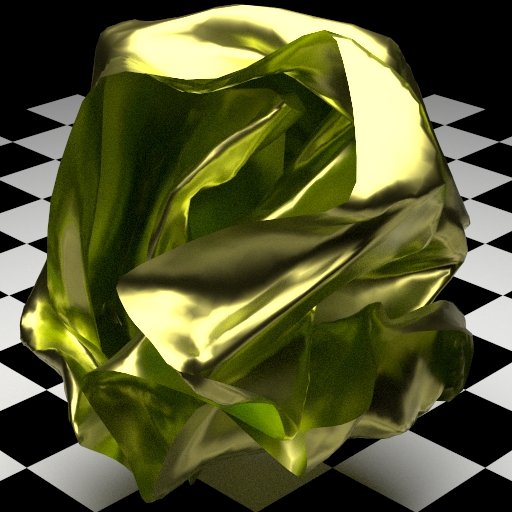} };
        \node[inner sep=0pt, right=1pt of A10] (A12) { \includegraphics[width=0.3\linewidth, frame]{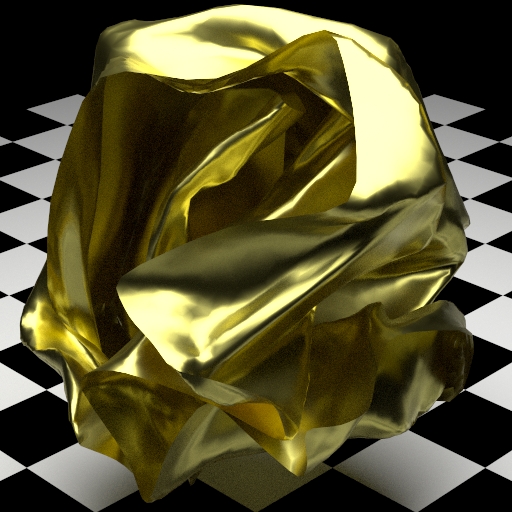} };
        \node[inner sep=0pt, right=1pt of A12] (A13) { \includegraphics[width=0.3\linewidth, frame]{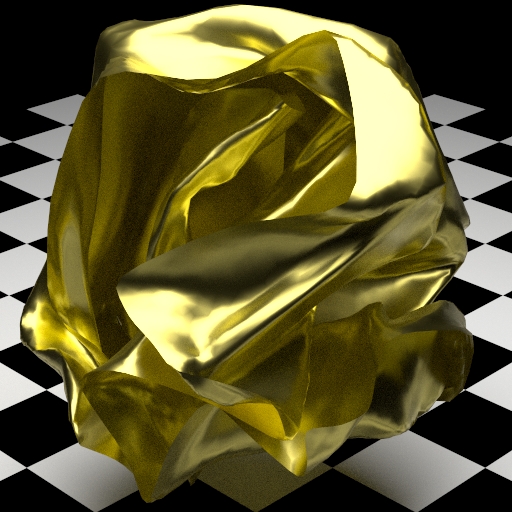} };
        \node[rotate=90, left=0pt of A10, anchor=south] (Ms) { {\textbf{Multiple Scattering}} };
    \end{tikzpicture}
    \vspace{-5pt}
    \caption{
    \textbf{Vathochromic reflectance.} 
    Multiple scattering has the same effect as a change in optical depth: it saturates the reflectance spectrum with a power law. Only this time, the exponents are integers. We use our vathochromic spectra as the $R_0$ component of a Schlick Fresnel in a microfacet model. While single scattering produces the same appearance, multiple scattering depicts a change in tint that we control (from green to yellow).
    \label{fig:results_vathochromism_paper}
    \vspace{-15pt}
    }
\end{figure}

\paragraph*{Vathochromic reflectance}
Vathochromic effects may also occur with reflectance spectra, due to inter-reflections on shiny (typically metallic) materials.
Figure~\ref{fig:results_vathochromism_paper} illustrates this effect on a crumpled paper model.
We use $K=9$ warped basis functions.
In this case, the target chromaticity $\mathbf{c}=[0.4, 0.43]^{\top}$ and luminance $F_y=0.59$ control the reflectance at normal incidence $R_0$ after a single scattering event.
We then use the parametrization shown in Figure~\ref{fig:results_usambara_unit_test}(left) to span the equivalence class of reflectance spectra, using Schlick's reflectance model~\cite{Schlick1994} to compute $R_0(\lambda)^d$ at normal incidence and the corresponding reflectance curves at discrete orders $d$ of inter-reflection.
This allows us to quickly find three different spectra that yield the same appearance in direct lighting, but exhibit the desired targeted hues in inter-reflections.

\subsection{Reproducing Metamerism}
\label{sec:apps-meta}

Our one-to-many sampler also permits the exploration of the space of metameric spectra.
Instead of directly using the PU to build a basis gamut in chromaticity space, we premultiply each element of the partition of unity with a target illuminant $I(\lambda)$:
\begin{align}
    B_k^I(\lambda) = B_k(\lambda) I(\lambda).
\end{align}
This defines a different gamut in chromaticity space per illuminant:
\begin{align}
    \mathbf{b}_k^I = \frac{\left[B_{k,X}^I, B_{k,Y}^I\right]^{\top}}{\vert B_k^I \vert}.
\end{align}

\begin{figure}[b]
    \centering
    \includegraphics[width=\linewidth]{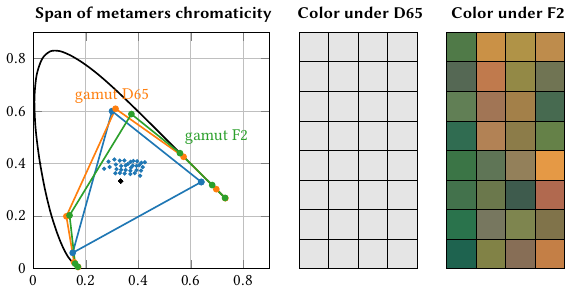}
    \vspace{-15pt}
    \caption{
    \textbf{Metameric palette.} We use our method to generate $32$ spectra that produce the same achromatic color $F_Y=0.8$ under a D65 (black dot and middle column) but provide a variety of colors under a F2 illuminant (blue dots and right column). We evaluate the accessible variability of metameric spectra under such a D65 constraint by random sampling. 
    \label{fig:results_metamerism_unit_test}
    \vspace{-25pt}
    }
\end{figure}

\begin{figure}[t]
    \centering
    \begin{tikzpicture}[font=\footnotesize]
        \node[inner sep=0pt] (D65) { \includegraphics[width=0.49\linewidth,frame, trim=0 0 0 80, clip]{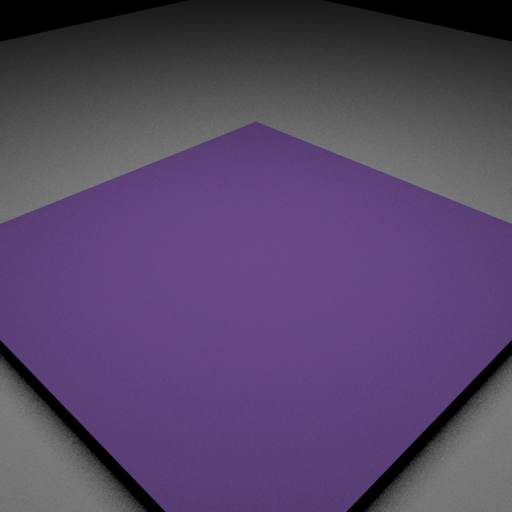} };
        \node[inner sep=0pt, right=2pt of D65] (F2) { \includegraphics[width=0.49\linewidth,frame, trim=0 0 0 80, clip]{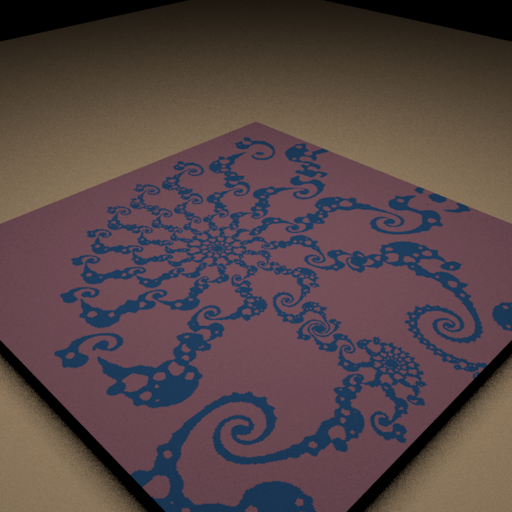} };
        \node[right=0pt of F2.north west, anchor=north west] (tex) { \includegraphics[width=0.15\linewidth, height=0.17\linewidth,frame,rotate=-90]{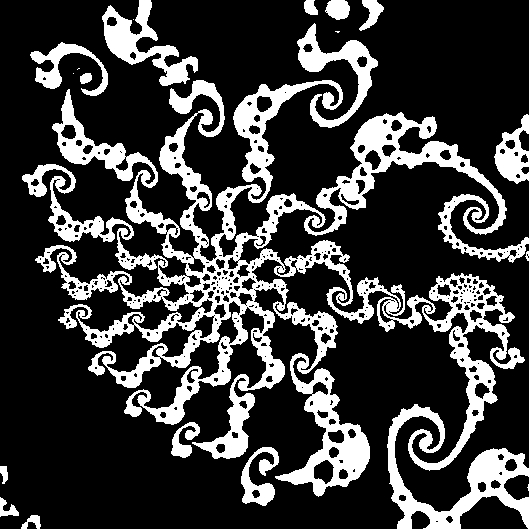} };
        \node[below=-13pt of D65, white] (D65text) {{\textbf{D65 illuminant}}};
        \node[below=-13pt of F2, white] (F2text) {{\textbf{F2 illuminant}}};
        \node[below=-5pt of tex, text=black] (textext) {\tiny \textbf texture};
        \node[below=-5pt of tex, text=white] (textext) {\tiny texture};
    \end{tikzpicture}
    \caption{\textbf{Metameric pattern.}
    We use a binary texture (inset, right) to drive the use of one of two metameric spectra that both appear purple under a $D65$ illuminant, but differ in color under a F2 illuminant.
    \label{fig:results_metamerism_julia}
    }
\end{figure}

\begin{figure}[b]
    \centering
    \begin{tikzpicture}[font=\footnotesize]
        \node[inner sep=0pt] (D65) { \includegraphics[width=0.49\linewidth,frame]{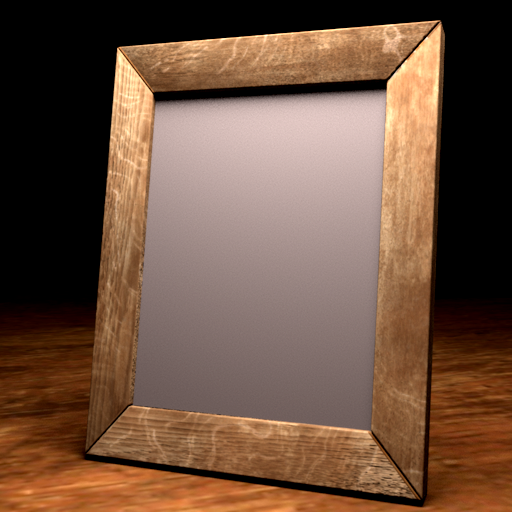} };
        \node[inner sep=0pt, right=2pt of D65] (F2) { \includegraphics[width=0.49\linewidth,frame]{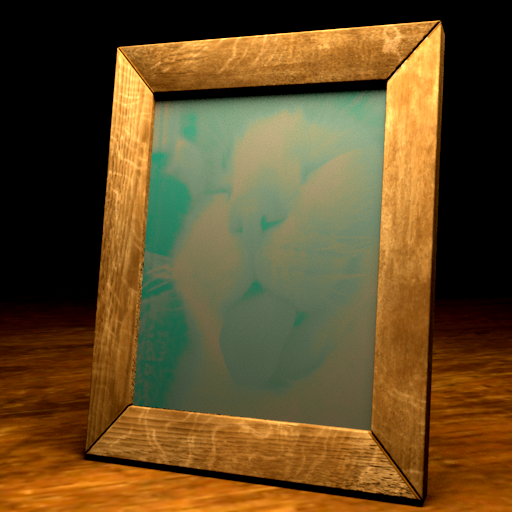} };
        \node[right=0pt of F2.north west, anchor=north west] (tex) { \includegraphics[width=0.15\linewidth, height=0.17\linewidth,frame]{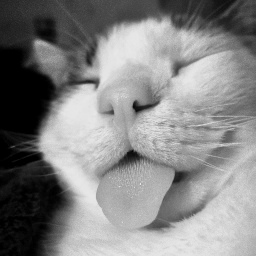} };
        \node[below=-13pt of D65, white] (D65text) {{\textbf{D65 illuminant}}};
        \node[below=-13pt of F2, white] (F2text) {{\textbf{F2 illuminant}}};
        \node[below=-5pt of tex, text=black] (textext) {\tiny  \textbf texture};
        \node[below=-5pt of tex, text=white, xshift=1pt] (textext) {\tiny  texture};
    \end{tikzpicture}
    \vspace{-10pt}
    \caption{
    \textbf{Metameric image.}
    In this example, we produce a dense set of metamer spectra to hide the photograph of a cat (inset, right) under D65 lighting.
    For each pixel, we blend 8 spectra using the target image's gray level. Blending factors are computed as a smooth Partition of Unity on $[0,1]$.
    \label{fig:results_metamerism_fermin}
    }
\end{figure}

\clearpage
\clearpage

Now, given a choice of illuminant -- say $D65$, we sample the equivalence class that achieves a target chromaticity $\mathbf{c}^{D65}$, yielding a set of vectors $\{\mathbf{w}^{D65}\}$ of basis coefficients.
When these vectors are used with the basis functions premultiplied by another illuminant -- say $F2$, they yield \emph{different} chromaticities since $\mathbf{b}_k^{F2} \neq \mathbf{b}_k^{D65}$.

\vspace{-5pt}
\paragraph*{Unit tests}
Metameric patches are shown in Figure~\ref{fig:results_metamerism_unit_test}, where a same achromatic color in $D65$ is shown to correspond to a variety of different chromaticities in $F2$, using $K=7$ \textit{non-warped} basis functions.
Each sample of the $D65$ equivalence class thus yields an element of the palette achievable through metamerism.
The greenish trend in the color palette is due to the choice of illuminant $F2$.

\vspace{-5pt}
\paragraph*{Hidden patterns and images}
If we assign two different spectra from a metameric palette to two different regions of a surface, we obtain the result shown in Figure~\ref{fig:results_metamerism_julia}, where the use $\mathbf{c}=[0.32, 0.25]^{\top}$ and $F_Y=0.8$.
Here the spectrum controls the spectral diffuse albedo of a Lambertian material.
The pattern is thus hidden under $D65$ illumination, but revealed under $F2$.
Note that for rendering, we use the original basis functions $B_k(\lambda)$, not the premultiplied ones.
A similar effect can be obtained with hidden images, as shown in Figure~\ref{fig:results_metamerism_fermin}.
Here we take a gray-level picture as input, and blend $8$ spectra of increasing luminance from the metameric palette of Figure~\ref{fig:results_metamerism_unit_test} to reproduce luminance gradients.
Compared to the tool of \cite{bergner2009tool}, our approach has two advantages: 1) their tool finds a single optimal metameric spectrum while we obtain a whole metameric palette; 2) they do not impose smoothness constraints on reflectance spectra while ours are smooth by design.

{
\subsection{Performance}}
\label{sec:perf}
As shown in the supplemental video, and although the prototype is implemented as a mono-threaded Python script, our upsampling method runs in real-time for artistic design. We measured timings for an increasing number of basis elements $K$ on an Intel i3-6100 CPU at $3.70$GHz. We report that our method runs at $4.9$ms for $K=5$, $13.1$ms for $K=7$, $19.3$ms for $K=9$, and $27$ms for $K=11$.

Note that using the Metameric Blacks construction~\cite{wyszecki1958evaluation} requires to track \change{}{a number} of constraints equals to twice the number of bins of a discretized spectra. In the literature, $31$ bins are commonly used. We measured that generating the convex hull for a binning of $21$ bins already takes $6.3$s on average per spectra when using the scipy's interface to the qhull library (with default parameters and double precision)\footnote{We do not count occurrences where the algorithm fails to find a solution.}.

%% file: tex/sec_conclusion.tex
\section{Discussion and future work}
\label{sec:discuss}

We have introduced a novel method to upsample a color to an equivalence class of spectra through a well-defined one-to-many mapping.
It provides another reason to move to spectral rendering besides the production of more photorealistic rendering results: the exploration of new visual effects as well as the imitation of those found in nature (gem stones, oils, etc). 
{}{We have focused in particular on the generation of vathochromic effects, both in transmission and reflection, generalizing the intriguing Usambara effect. 
We have also shown how our approach applies to metameric effects.
We now discuss its specificities.}

\begin{figure}[h]
    \center
    \includegraphics[width=\linewidth]{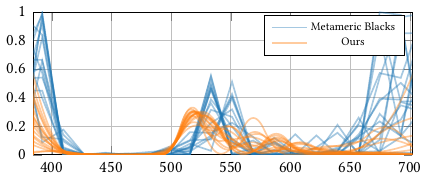}
    \vspace{-20pt}
    \caption{\change{}{\textbf{Comparison with Metameric Blacks (MB).}
        Both our method and Metameric Blacks permit to sample all possible metamers for a target color (here $\mathbf{C} = [0.05, 0.1, 0.01]$) in their respective equivalence class. 
        However, MB uses a binned representation (here $21$ bins) which yields unrealistic reflectance spectra.
        }
        \label{fig:cmp_metamer_blacks}
    }
\end{figure}

{}{
\paragraph*{Differences with Metameric Blacks}
The main difference with previous work in optics resides in the way reflectance spectra are represented. 
Methods that rely on discretized spectra (e.g.,~\cite{wyszecki1958evaluation,cohen1982metameric}) require a small number of spectral bins to be computationally tractable, as discussed in Sections~\ref{sec:prev} and~\ref{sec:perf} \change{}{and result in unrealistic spectra (Figure~\ref
{fig:cmp_metamer_blacks})}.
Another difference lies in the ease with which metameric sets may be explored. 
In our approach, an artist can quickly pick an achievable chromaticity (in the basis gamut), for which a maximum achievable luminance is readily provided through $\mathbf{\overline{w}}$ (Equation~\ref{eqn:maxLum-weight}).
In contrast, with the metameric blacks approach, when a target color results in an empty metameric set, users have to go through trial and error to find a color for which at least one spectrum exists.

An alternative is to rely on measured spectra, such as in the work of Finlayson and Morovic~\cite{finlayson2005metamer}.
Similar to Schmitt~\cite{schmitt1976method}, they reconstruct metamers using barycentric coordinates in the space of spectra. That is, from a set of $K$ measured spectra $s_k(\lambda)$ they reconstruct $r(\lambda) = \sum w_k s_k(\lambda)$ where $w_k$ are positive weights with $\sum w_k = 1$ (Equation~(27) in their paper). 
This imposes that $r(\lambda)$ is in the convex hull of the $s_k(\lambda)$. Our method only imposes validity constraints, $w_k \in [0,1]$. 
Therefore, we can always reconstruct perfect blacks ($r(\lambda) = 0, \, \forall \lambda$), perfect whites ($r(\lambda) = 1, \, \forall \lambda$), and achieve any target luminance $Y$.
On the other hand, their method trivially yields physically-realistic spectra, whereas ours is more adapted to artistic exploration. 
}

\begin{figure}[h]
    \centering
    \begin{tikzpicture}[font=\tiny]
        \begin{scope}
            \node[inner sep=0pt] (B) { \includegraphics[width=0.36\linewidth, frame, trim=0 64 0 64, clip]{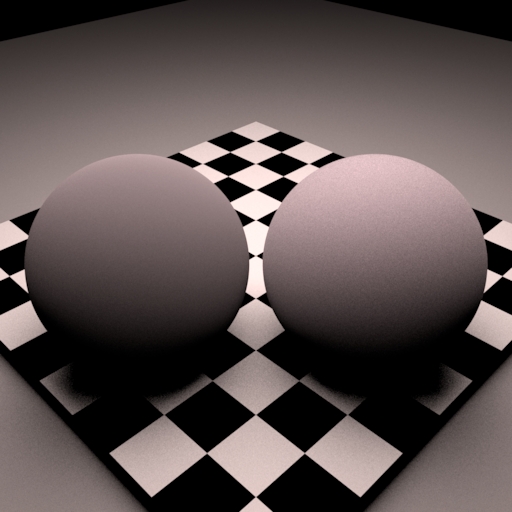} };
            \draw[color=red] {($(B.center) + (0.5,0.05)$)} rectangle +(0.4, 0.3);
            \node[inner sep=0pt, anchor=south, above=-9pt of B, white] { \textbf{{Target $Y=0.6$}} };
            \begin{scope}[shift={($(B.north east)+(0.01,0.0)$)}, anchor=north west]
                \node[inner sep=0pt] (A00) { \includegraphics[width=0.102\linewidth, cfbox=red 0.5pt 0pt, trim=354 232 64 200, clip]{figures/limitations/scene_test_a=99_06_opt_0000.jpg} };
                \node[inner sep=0pt, below=0pt of A00] (A01) { \includegraphics[width=0.102\linewidth, frame, trim=354 232 64 200, clip]{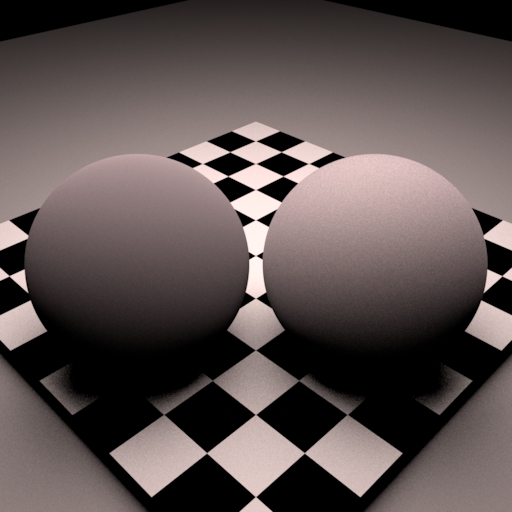} };
                \node[inner sep=0pt, below=0pt of A01] (A02) { \includegraphics[width=0.102\linewidth, frame, trim=354 232 64 200, clip]{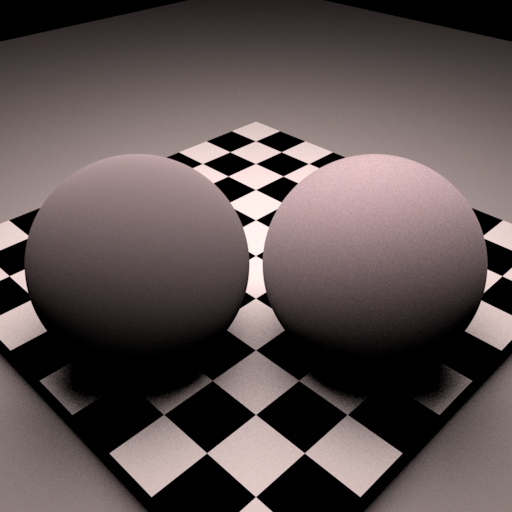} };
            \end{scope}
        \end{scope}
        \begin{scope}[xshift=0.5\linewidth]
            \node[inner sep=0pt] (B) { \includegraphics[width=0.36\linewidth, frame, trim=0 64 0 64, clip]{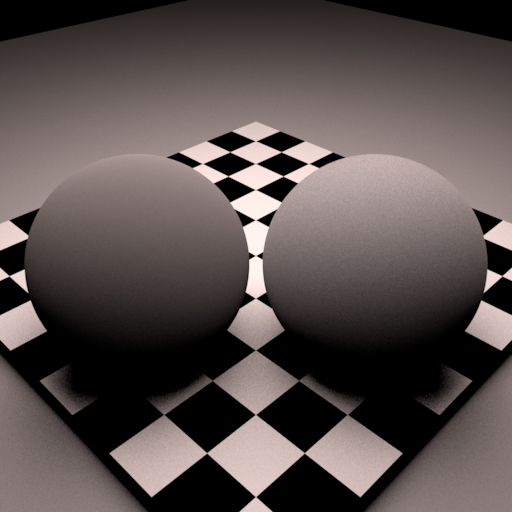} };
            \draw[color=red] {($(B.center) + (0.5,0.05)$)} rectangle +(0.4, 0.3);
            \node[inner sep=0pt, anchor=south, above=-9pt of B, white] { \textbf{{Target $Y=0.5$}} };
            \begin{scope}[shift={($(B.north east)+(0.01,0.0)$)}, anchor=north west]
                \node[inner sep=0pt] (A00) { \includegraphics[width=0.102\linewidth, cfbox=red 0.5pt 0pt, trim=354 232 64 200, clip]{figures/limitations/scene_test_a=99_05_opt_0001.jpg} };
                \node[inner sep=0pt, below=0pt of A00] (A01) { \includegraphics[width=0.102\linewidth, frame, trim=354 232 64 200, clip]{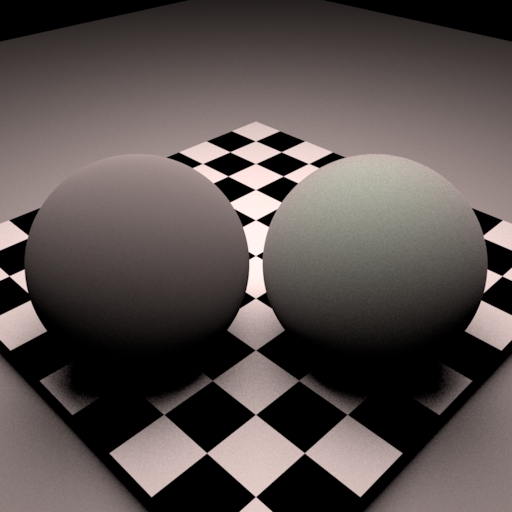} };
                \node[inner sep=0pt, below=0pt of A01] (A02) { \includegraphics[width=0.102\linewidth, frame, trim=354 232 64 200, clip]{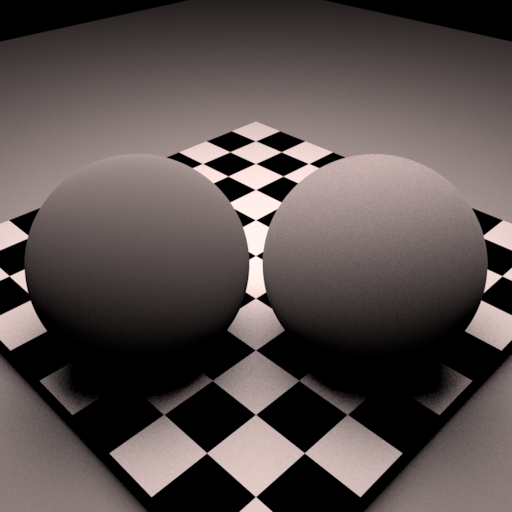} };
            \end{scope}
        \end{scope}
    \end{tikzpicture}
    \caption{
    \textbf{Limitations.} 
    We use vathochromic spectra for Fresnel reflectance to control the color of multiple scattering within microfacet models~{\protect\cite{heitz2016multiple}}. 
    The left and right spheres are rendered using single and multiple scattering respectively, with a roughness $\alpha=0.99$. 
    The insets show three samples from the equivalence class.
    Above some luminance value (left), it is difficult to achieve a variety of colors in the generated spectra. 
    Reducing the target luminance (right) produces slightly more diverse spectra.
    \label{fig:limitations}
    \vspace{-10pt}
    }
\end{figure}

\paragraph*{Limitations}
An inherent limitation of spectral asset creation, already pointed out by MacAdam~\shortcite{MacAdam35a,MacAdam35b}, is that one needs to trade saturation for luminance.
Indeed, saturated spectra necessarily have narrow bands, and our approach is no different  in this respect.
This limitation might explain the difficulty to create visually-noticeable vathochromic effects in microfacet models to control the color of the multiple scattering term (see Figure~\ref{fig:limitations}).

A direction of improvement for our method lies in the design of techniques to navigate through equivalence classes. 
In particular, it would be useful to give an analytical description of bounds imposed by the target luminance $F_Y$ (i.e., the boundary between opaque and transparent points in Figure~\ref{fig:explore_bary}(middle)).

Last, even though the spectra generated by our approach are physically-plausible, they are not physically-realistic by design.
Real spectra obey physical rules of their own.
For instance, the real and imaginary parts of refractive indices are bound by the Kramers-Kronig relations.
It would thus be interesting to establish connections with physical models of spectra.
In this respect, having a large equivalence class from which to pick  spectra closest to physically-realistic ones could be an advantage.

\paragraph*{Future work}
Our method could be used to reproduce and study a number of interesting optical phenomena.
The Alexandrite effect, an instance of metamerism, is one famous example: with our approach, we could investigate whether other spectra could potentially create similar effects.
Interesting applications could be found in ecology, where the illuminant plays a crucial role in defining habitats.
For instance, we could study how families of spectra are affected by lighting at different depths under water or under a dense foliage.
We would also like to explore the extension of vathochromism to take into account fluorescence effects, which abound in nature.

Finally, we have only considered normal human color vision through the use of CIE sensitivity functions.
A captivating direction of future work would be to experiment with sensitivity functions adapted to color blindness, or even to animal vision.